\documentstyle[aps,pre,multicol,psfig]{revtex}

\draft

\begin{document}

\title{Universality classes in the random-storage sandpile model}

\author{Alexei V\'azquez$^1$ and Oscar Sotolongo-Costa$^{1,2}$}

\address{$^1$ Department of Theoretical Physics, Faculty of
Physics, Havana University, Havana 10400, Cuba}

\address{$^2$ LICTDI, Faculty of Sciences, UNED, Madrid 28080,
Spain}

\maketitle

\begin{abstract}

The avalanche statistics in a stochastic sandpile model where
toppling takes place with a probability $p$ is investigated. The
limiting case $p=1$ corresponds to the Bak-Tang-Wiesenfeld (BTW)
model with deterministic toppling rule. Based on the moment
analysis of the distribution of avalanche sizes we conclude that
for $0<p<p_c$ the model belongs to the DP universality class
while for $p_c<p<1$ it belongs to the BTW universality class,
where $p_c$ is identified with the critical probability for
directed percolation in the corresponding lattice.

\end{abstract}


\begin{multicols}{2}


Sandpile automata were proposed as a paradigm of self-organized
critical (SOC) phenomena \cite{bak}. These simple models capture
its essential dynamics, which takes place in the form of
avalanches of all sizes. At the early state of SOC theory it was
believed that the critical state of sandpile automata is
insensitive to changes in model parameters, however some recent
works contradict this statement. For instance, Vespignani and
Zapperi \cite{vespignani} have shown that driving and
dissipation rates actually act as control parameters,
criticality is obtained after fine tuning of these fields. On
the other hand, we have recently shown that a class of models
with stochastic rules display a transition from SOC to directed
percolation (DP) with increasing the degree of stochasticity
\cite{vazquez}. Nevertheless, before we make our final
conclusion, we have to investigate if the original
Bak-Tang-Wiesenfeld (BTW) sandpile automaton and these modified
models belong to the same universality class, otherwise they
would just be different models. One may thus ask: do
deterministic and stochastic sandpile models belong to the same
universality class?

The BTW automaton is defined on $d$-dimensional hypercubic
lattice of linear size $L$. On each site $i$ an integer variable
$z_i$ is defined, which we call energy following
\cite{vespignani}. Energy is added to the system by selecting a
site at random and increasing its energy by one, i.e.
$z_i\rightarrow z_1+1$. The energy addition continues until one
of the sites reaches or exceeds a threshold $z_c=2d$, then this
site topples transferring energy to its neighbors. Toppling is
defined by the set of rules $z_i\rightarrow z_i-z_c$ and
$z_j\rightarrow z_j+1$ at each of the $j$ nearest neighbors. The
toppling at one site may induce an avalanche of toppling events.
After the avalanche has stopped we re-start adding energy to the
system. Open boundary conditions are assumed. The toppling rules
of this model are deterministic and the only randomness is
introduced through the random energy addition.

May be the simplest stochastic sandpile model is the Manna or
$d$-state model \cite{manna}. In this case $z_c=d<2d$ and,
therefore, only $d$ of the $2d$ neighbors will receive energy
after toppling. These $d$ neighbors are selected at random.
Real-space renormalization group approach \cite{pietronero}
suggests that the BTW and Manna models belongs to the same
universality class. However, there is not complete agreement in
numerical simulations. Early large scale simulations of the
Manna \cite{manna} and BTW models \cite{grassberger} in two
dimensions, show that the avalanche distributions are described
by the same exponents for the power law decay and the scaling of
the cutoffs. These results where supported by more recent
estimates of the avalanche exponents by L\"ubeck and Usadel
\cite{lubeck}. On the contrary, Ben Hur and Biham \cite{ben}
analyzed the scaling of conditional expectation values of
various quantities, obtaining significant differences in the
exponents for the two models.  However, Chessa {\em et al} have
recently shown \cite{chessa} that the method of conditional
expectation values, introduced by Ben Hur and Biham, is
systematically biased by non-universal corrections and,
therefore, does not provide indications on universality classes.
Moreover, according to the large scale simulations performed by
Chessa {\em at al} \cite{chessa}, the BTW and Manna models
belong to the same universality class.

In the Manna model the randomness appears in the energy transfer
after toppling, but the condition for toppling is deterministic.
Motivated on a directed model by T\'adic and Dhar \cite{tadic}
we have proposed a different model, where sites topples when
$z\geq z_c=2d$ but with probability $p$ \cite{vazquez}. In this
case, as in the BTW model, each of the $2d$ neighbors receives
one energy grain but the condition for toppling is stochastic.
In the particular case $p=1$ one recovers the BTW model. For
$p<1$ sites accumulate a random amount of energy, random storage
model (RSM) \cite{vespi}. From mean-field theory and numerical
simulations in one dimension \cite{vazquez} we obtain that the
RSM self-organizes into a critical state for $p_c<p<1$, while it
is subcritical for $p<p_c$. The subcritical state is found to be
similar to DP and $p_c$ is identified with the critical
threshold for DP. The correlation length exponents are identical
to those of DP but other exponents result different due to open
boundary conditions. However, in that occasion our analysis was
limited to a small region close to $p_c$ and simulations were
performed only in $d=1$. Thus, we could not provide any
indication about the universality classes.

In the present work we extend the numerical simulations of the
RSM to the whole range of $p$, in one and two dimensions. To
determine the avalanche size and duration exponents we use
moment analysis, a technique previously introduced by de Menech
{\em at al} \cite{menech} to obtain the scaling exponents for
the BTW model.  techniques. The numerical evidence suggests that
the RSM and the BTW model belong to different universality
classes. Moreover, for $p<p_c$ we corroborate that the model is
similar to DP.

The MF theory of the RSM \cite{vazquez} reveals that there is a
critical probability $p_c$ below which the system is
subcritical. In this regime $p$ is small and sites accumulates a
large amount of grains. Let us assume that all sites have
heights above the critical threshold $z_c$. Then when a new
grain is added at certain site this site will topples with
probability $p$. If it topples then each of its nearest
neighbors will also topple with probability $p$ and so on. This
picture is equivalent to a site directed percolation problem,
where sites become active with probability $p$ if one of its
neighbors was active in the previous step. Hence the correlation
length is given by $\xi\sim(p_c-p)^{-\nu}$, where $p_c$ is the
critical threshold for site directed percolation in the
corresponding lattice and $\nu$ is the spatial correlation
length exponent. Now, if $p\ll p_c$ in such a way that $\xi\ll
L$, where $L$ is the system size, then an avalanche starting
from a bulk site will never reach the boundary and, therefore,
no grain will be dissipated. Hence, the average height will
increase in time because dissipation will no balance the grain
addition from the external field. After a time long enough most
sites will have heights above $z_c$, supporting our starting
assumption. The dynamical state below $p_c$ is thus equivalent
to DP and $p_c$ is identified with the site DP threshold in the
corresponding lattice. Above $p_c$ and for $z>z_c$ at all sites
the system will still be equivalent (at least for small times)
to DP and an infinite avalanche will be generated. However, such
state is not stable because the infinite avalanche will reach
the boundary leading to dissipation of grains and hence the
decrease of the column heights up to the avalanche has stopped.
Above $p_c$ the stationary state is no more equivalent to DP.
The balance between the  grains added by the external field and
dissipation at the boundary leads to a SOC state. Our task will
be to determine if this SOC state belongs to the same
universality class of the BTW model.

Let $s$ be the avalanche size and $T$ its duration. $s$ is the
number of toppling events in an avalanche. $T$ is the number of
steps required to obtain a stable configuration starting from
the initial site with $z\geq z_c$, which triggered the
avalanche, taking into account that on each step all sites are
updated in parallel following the toppling rule. Both, $s$ and
$T$, are random variables. Their distributions are given by,
assuming scaling,
\begin{equation}
P_x(x)=x^{-\tau_x}f_x(x/x_c),
\label{eq:1}
\end{equation}
where $x$ is $s$ or $T$, $s_c$ and $T_c$ are the cutoff
avalanche size and duration, and $f_x$ is a cutoff function. The
cutoffs scale with the characteristic length of the system $\xi$
according to
\begin{equation}
x_c\sim \xi^{\beta_x},
\label{eq:2}
\end{equation}
where $\beta_s=D$, $\beta_t=z$, $D$ is the avalanche dimension
and $z$ the dynamic scaling exponent. In the SOC state $\xi\sim
L$ and, therefore, the cutoffs scale with system size. On the
contrary, in the subcritical state $\xi\sim(p_c-p)^{-\nu}$ and,
hence, the cutoffs diverges when $p$ approaches $p_c$. The
scaling exponents $\tau_x$ and $\beta_x$ are not all
independent. From the identity $P(s)ds=P(T)dt$ one obtains
\begin{equation}
(\tau_s-1)D=(\tau_t-1)z.
\label{eq:2a}
\end{equation}
On the other hand, in the SOC state ($p>p_c$) the mean avalanche
size should scale as $\langle s\rangle\sim L^2$ and, therefore,
\begin{equation}
(2-\tau_s)D=2.
\label{eq:2b}
\end{equation}
These scaling relations are useful to test the reliability of
the numerical estimates.

The purpose of present numerical simulations is to determine the
exponents $\tau_s$, $\tau_t$, $D$ and $z$. We are going to take
up this task using as fundamental technique the moment analysis
\cite{menech}.  The $q$ moment is given by
\begin{equation}
\langle x^q\rangle = \int dx P(x)x^q \sim \xi^{\sigma_x(q)},
\label{eq:3}
\end{equation}
where
\begin{equation}
\sigma_x(q)=\beta_x(1-\tau_x)+\beta_x q.
\label{eq:4}
\end{equation}
The last equivalence in eq. (\ref{eq:3}) is not valid for small
values of $q$. For small $q$ the integral depends in the
functional form of $P(x)$ in the whole range of $x$, while
scaling assumptions are in general not valid for small $x$. The
extreme case is $q=0$, normalization imposes $\sigma_x(0)=0$
and, therefore, eq. (\ref{eq:4}) is not valid. But, for
$q\geq\tau_x-1$ large $x$ dominates leading to the last
equivalence in eq. (\ref{eq:3}).

After computing the moments one can obtain $\sigma_x(q)$ from a
liner fit to the log-log plot of $\langle x^q\rangle$ vs. $\xi$.
Then one can obtain $\tau_x$ and $\beta_x$ from a linear fit to
the straight part of the plot $\sigma_x(q)$ vs. $q$. Above $p_c$
we have $\xi\sim L$, but it is a function of $p$ below. To
compute the correlation lenght below $p_c$ we use the following
expression \cite{vazquez}
\begin{equation}
\xi^2\sim \frac{\sum_{t=0}^\infty\sum_{i=0}^L(i-i_0)^2\rho_{ai}}
{\sum_{t=0}^\infty\sum_{i=0}^L\rho_{ai}},
\label{eq:5}
\end{equation}
where $i_0$ is the position of the initial active ($z\geq z_c$)
site, $t$ is the number of steps measured in the time scale of
the avalanche and $\rho_{ai}=1$ ($\rho_{ai}=0$) in active
(inactive) sites. Moreover, from the log-log plot of $\xi$ vs.
$p_c-p$ one can obtain a numerical estimate of $p_c$ and $\nu$.

\begin{figure}\narrowtext
\centerline{\psfig{figure=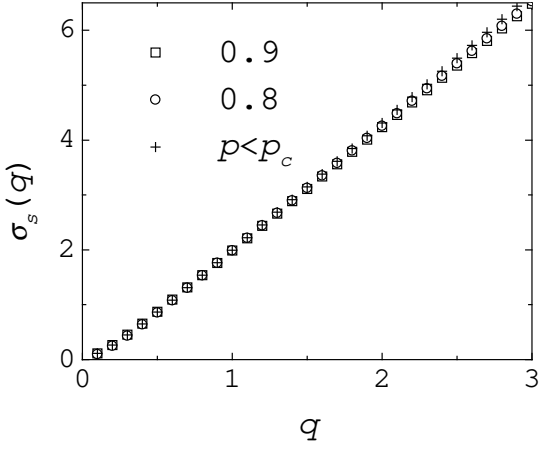,width=3in}}
\caption{Plot of $\sigma_s(q)$ in $d=1$ for different values of
$p$. Data above $p_c=0.707(2)$ where obtained using lattice
sizes $L=80$, 160, 320 and 640. Data below $p_c$ where obtained
using probabilities $p=0.670$, 0.688, 0.696 and 0.7000.}
\label{fig:1}
\end{figure}

\begin{figure}\narrowtext
\centerline{\psfig{figure=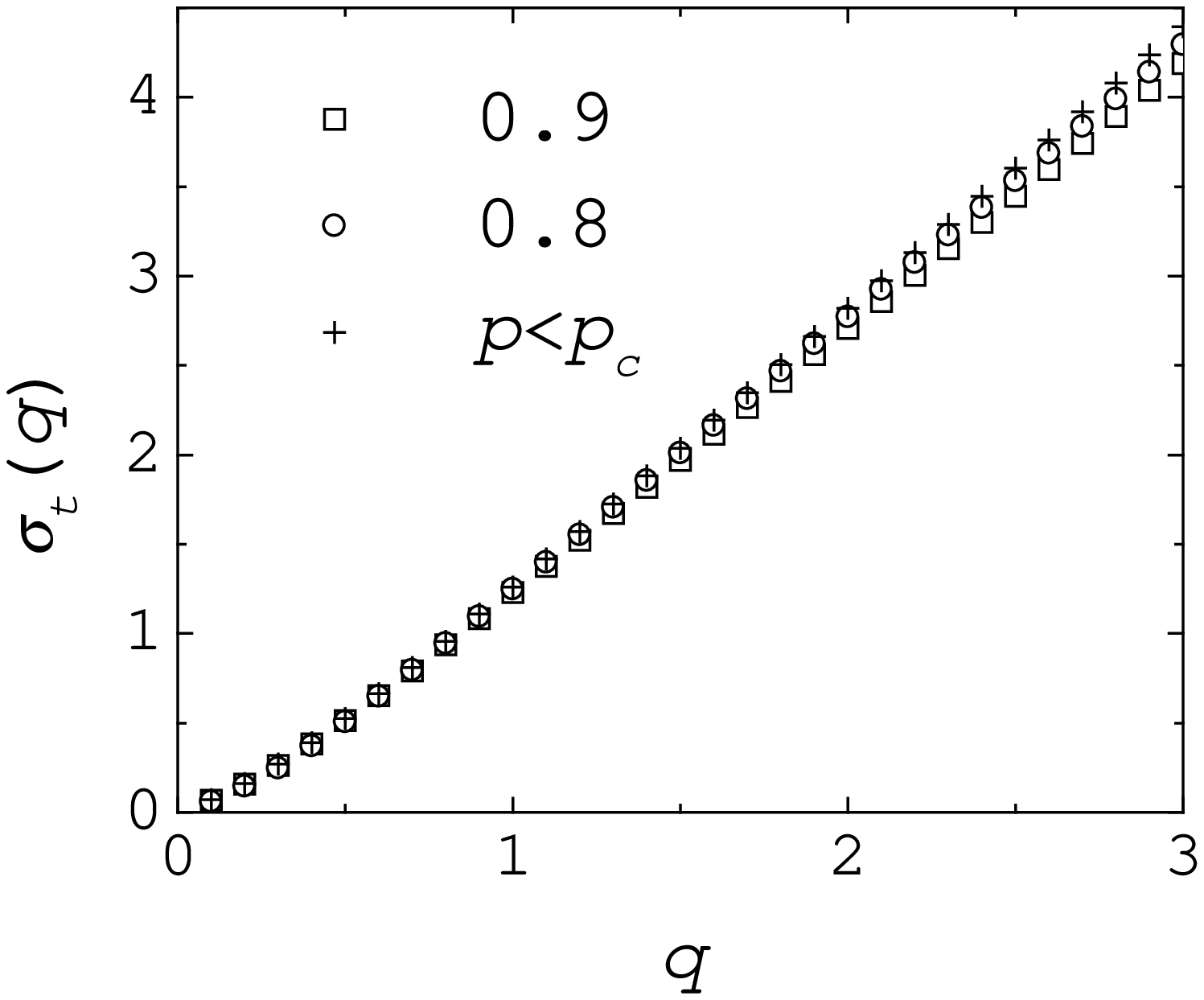,width=3in}}
\caption{Plot of $\sigma_t(q)$ in $d=1$. The lattice sites and
probabilities used are the same as described in the caption of
fig. \ref{fig:1}.}
\label{fig:2}
\end{figure}

$d=1$: The BTW model ($p=1$) in one dimension exhibits trivial critical
behavior \cite{bak}. No power law behavior is observed in the
avalanche size and duration distributions, however one can
compute the exponents $D$ and $z$ from the scaling of the
moments with system size, resulting $D\approx2$ and $z\approx1$.
Using this values and the scaling relations in eqs (\ref{eq:2a})
and (\ref{eq:2b}) one could obtain the exponents $\tau_s$ and
$\tau_t$ ($\tau_s=\tau_t=1$) however these scaling laws are not
valid in this case because the distributions of avalanche size
and duration do not follow the scaling law in eq. (\ref{eq:1}).
On the contrary, the RSM in one dimension has non trivial
behavior. The $q$ dependence of $\sigma_s$ and $\sigma_t$, for
different values of $p<1$, is shown in figs. \ref{fig:1} and
\ref{fig:2}, respectively. The numerical estimates of the
scaling exponents are given in table \ref{tab:1}. For
$p_c<p\leq0.8$ we observe that $\sigma_s(q)$, $D$, $\tau_s$ and
$\tau_t$ are practically insensitive to changes in $p$, but
systematic deviations are observed for $\sigma_t(q)$ and $z$.
For $p=0.9$ the scaling exponents are between those for $p=1$
and $p=0.8$. On the other hand, using the numerical data below
$p_c$, we have obtained $p_c=0.707\pm0.002$, $\nu=1.07\pm0.03$
and $z=1.57\pm0.02$ which are, within the numerical error,
identical to the series expansion estimates for site DP in a
square lattice \cite{jenssen}. Moreover, the exponents
$\delta=\tau_t-1$ and $z$ are consistent with previous numerical
simulations \cite{vazquez}. We have also carry out finite size
scaling of the distributions of avalanche size and duration. In
all cases, including the BTW limit, we observe a good data
collapse and the obtained scaling exponents are in agreement
with those obtained from the moment analysis. Moreover, within
the numerical error, the scaling relations in eqs. (\ref{eq:2a})
and (\ref{eq:2b} are satisfied.

\begin{figure}\narrowtext
\centerline{\psfig{figure=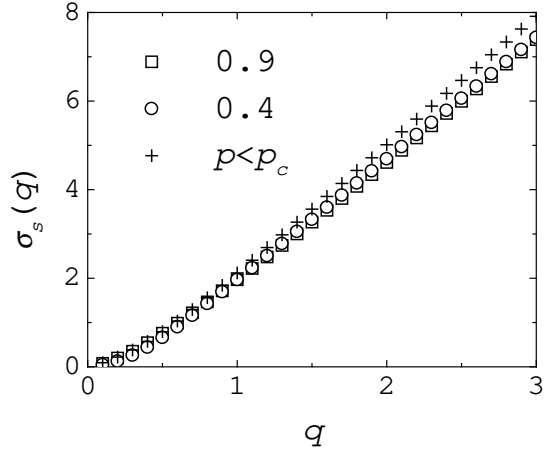,width=3in}}
\caption{Plot of $\sigma_s(q)$ in $d=2$ for different values of
$p$. Data above $p_c=0.344(1)$ where obtained using lattice
sizes $L=64$, 128, 256 and 512. Data below $p_c$ where obtained
using probabilities $p=0.26$, 0.30, 0.33 and 0.34.}
\label{fig:3}
\end{figure}

\begin{figure}\narrowtext
\centerline{\psfig{figure=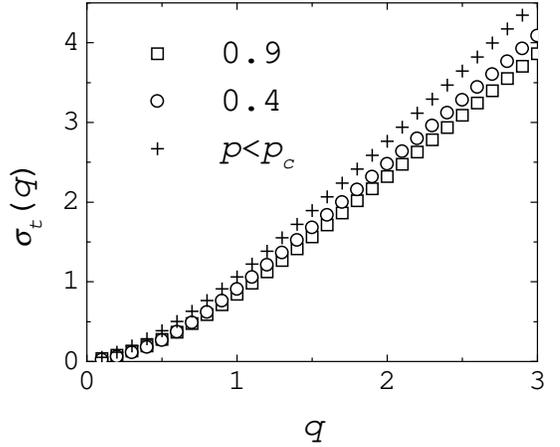,width=3in}}
\caption{Plot of $\sigma_t(q)$ in $d=2$. The lattice sites and
probabilities used are the same as described in the caption of
fig. \ref{fig:3}.}
\label{fig:4}
\end{figure}

$d=2$: The BTW model has nontrivial exponents. However, the data
collapse was not compatible with the scaling assumption in eq.
(\ref{eq:1}). The corrections to scaling in the BTW has been
found to be very strong \cite{lubeck,chessa}, making necessary
the use of very large lattice sizes to obtain accurate estimates
of the scaling exponents. The largest lattice size used in our
simulations, $L=512$, seems to be not large enough. Using
lattice sizes ranging from $L=512$ to $L=2048$ Chessa {\em et
al} \cite{chessa} have obtained a good finite size scaling for
the distribution of avalanche size, but have not for the
distribution of avalanche duration. We thus rule out the
possibility of determine the BTW exponents in two dimensions
with such small lattice sizes. Instead of that, we are going to
use the numerical estimates by Chessa {\em et al} \cite{chessa}
and L\"ubeck and Usadel \cite{lubeck}. On the contrary, the RSM
displays good finite size scaling for the lattice sizes we have
used. Moreover, the scaling exponents obtained from the finite
size scaling are in agreement with those obtained from the
moment analysis. The $q$ dependence of $\sigma_s(q)$ and
$\sigma_t(q)$ is shown in figs. \ref{fig:3} and \ref{fig:4},
respectively. The numerical estimates of the scaling exponents
for different values of $p<1$ are given in tab. \ref{tab:2},
toguether with the reports by Chessa {\em et al} \cite{chessa}
and L\"ubeck and Usadel \cite{lubeck} for the BTW model.
$\sigma_s(q)$, $D$, $\tau_s$ and $\tau_t$ are practically
insensitive to changes in $p$ above $p_c$, even considering the
deterministic limit. On the contrary, $\sigma_t(q)$ and $z$
suffer from systematic deviations with changing $p$. In this
case we must be more careful because the finite size effects
have stronger influence on the distribution of avalanche
duration. For instance, for the largest lattice size used
$L=512$ the distribution of avalanche sizes cover about six
decades while the distribution of avalanche durations cover less
than five decades. To obtain more precise determination of the
dynamic scaling exponents we must increase system size. In the
mean time, the scaling exponents of the distribution of
avalanche sizes indicates that the RSM in the range $p_c<p<1$
belongs to the same universality class of the BTW model, which
correspond with the limit $p=1$. On the other hand, below $p_c$
we have obtained $p_c=0.344\pm0.001$, $\nu=0.728\pm0.002$ and
$z=1.76\pm0.02$ which are, within the numerical error, identical
to the numerical estimates for site DP in a BCC lattice
\cite{grassberger1}. In this case the difference between the
exponents below and above $p_c$ is significant, showing that the
RSM above and below $p_c$ belongs to different universality
classes.

The numerical simulations in $d=2$ corroborate that the RSM is
similar to DP below $p_c$. As in $d=1$, the exponents $\nu$ and
$z$ and the critical probability $p_c$ are identical to the DP
values in the corresponding lattice. On the other hand, the
difference between the exponents $D$ and $z$, obtained using the
data below $p_c$ and those obtained above but close to $p_c$, is
far form being contained within the error bars. The model above
and below $p_c$ belong to different universality classes. Below
$p_c$ it is DP with open boundaries and random initial seed,
while above $p_c$ we will continue using the term RSM.

The numerical evidence obtained for the avalanche size
distribution indicates that in $d=2$ the RSM belong to the
universality class of the BTW model. The scaling exponents
$\tau_s$, $\tau_t$ and $D$ are practically independent of $p$ in
the SOC regime $p_c<p\leq1$, however, $z$ shows a strong $p$
dependence which may be attributed to finite size effects. On
the contrary in $d=1$ the distribution of avalanche sizes and
duration for the BTW model display trivial behavior while in the
RSM they satisfy the scaling hypothesis. In larger dimensions
$d>2$ we expect that the SOC regime of the RSM belongs to the
BTW universality class as in $d=2$ .

In summary the RSM has three different regimes. I: $0<p<p_c$
similar to DP, II: $p_c<p<1$ where the toppling rule are still
stochastic but the system is in a SOC state and III: $p=1$ (BTW)
the toppling rules are deterministic. Based on the moment
analysis of the distribution of avalanche sizes we conclude that
for $0<p<p_c$ the model belongs to the DP universality class
while for $p_c<p<1$ it belongs to the BTW universality class.

We thanks A. Vespignani for bringing up our attention on
two-dimensional simulations. This work was partially supported
by the {\em Alma Mater} prize, given by the University of
Havana.


\begin{table}\narrowtext
\begin{tabular}{lllllll}
$p$ & $D$ & $z$ & $\tau_s$ & $\tau_t$\\ \hline
1 & 2 & 1 & &\\
0.9 & 2.25(1) &	1.48(2) & 1.12(1) & 1.17(2)\\
0.8 & 2.27(1) &	1.52(2) & 1.12(1) & 1.18(2)\\
0.708 &	2.27(1) & 1.54(2) & 1.13(1) & 1.18(2)\\
$p<p_c$ & 2.34(1) & 1.57(2) & 1.16(1) & 1.20(2)
\end{tabular}
\caption{Scaling exponents in $d=1$ for different values of
$p$.}
\label{tab:1}
\end{table}

\begin{table}\narrowtext
\begin{tabular}{lllllll}
$p$ & $D$ & $z$ & $\tau_s$ & $\tau_t$\\ \hline
1   & 2.73(2)\cite{chessa} & 1.52(2)\cite{chessa} 
& 1.293\cite{lubeck}   
& 1.480\cite{lubeck}\\
0.9 & 2.74(1) &	1.53(2) & 1.28(1) & 1.48(2)\\
0.8 & 2.75(1) &	1.54(2) & 1.29(1) & 1.48(2)\\
0.6 & 2.74(1) &	1.58* & 1.29(1) & 1.47*\\
0.4 & 2.74(1) &	1.61* & 1.28(1) & 1.46*\\
$p<p_c$ & 2.90(1) & 1.76(2) & 1.27(1) &	1.43(2)
\end{tabular}
\caption{Scaling exponents in $d=2$ for different values of
$p$. * These exponents may be affected by finite size corrections.}
\label{tab:2}
\end{table}

\vskip 1in

\vskip 1in

\end{multicols}


\begin{thebibliography}{50}

\bibitem{bak} P. Bak, C. Tang, and K. Wiesenfeld, Phys. Rev.
Lett. {\bf 59}, 381 (1987); Phys. Rev. A {\bf 38}, 364 (1988).

\bibitem{vespignani} A. Vespignani and S. Zapperi, Phys. Rev.
Lett. {\bf 78}, 4793 (1997); Phys. Rev. E {\bf 57}, 6345 (1998).

\bibitem{vazquez} A. V\'azquez and O. Sotolongo-Costa,
cond-mat/9810408 (submitted to J. Phys. A).

\bibitem{manna} S. S. Manna, Physica A {\bf 179}, 249 (1991).

\bibitem{pietronero} L. Pietronero, A. Vespignani, S. Zapperi,
Phys. Rev. Lett. {\bf 57}, 1690 (1994); A. Vespignani, S.
Zapperi, and L. Pietronero, Phys. Rev. E {\bf 51}, 1711 (1995).

\bibitem{grassberger} P. Grassberger and S. S. Manna, J. Phys.
(France) {\bf 51}, 1077 (1990); S. S. Manna, J. Stat. Phys.
{59}, 509 (1990); Physica A {\bf 179}, 249 (1991).

\bibitem{lubeck} S. L\"ubeck and K. Usadel, Phys. Rev. E {\bf
55}, 4095 (1997).

\bibitem{ben} A. Ben-Hur and O. Biham, Phys. Rev. E {\bf 53},
R1317 (1996).

\bibitem{chessa} A. Chessa, H. E. Stanley, A. Vespignani, and S.
Zapperi, cond-mat/9808263.

\bibitem{tadic} B. T\'adic and D. Dhar, Phys. Rev. Lett. {\bf
79}, 1519 (1997).

\bibitem{vespi} A. Vespignani, personal communication.

\bibitem{menech} M. de Menech, A. L. Stella, and C. Tebaldi, 
Phys. Rev. E {\bf 58}, R2677 (1998).

\bibitem{jenssen} I. Jenssen, J. Phys. A {\bf 29}, 7013 (1996).

\bibitem{grassberger1} P. Grassberger, J. Phys. A {\bf 22}, 3673
(1989).

\end{thebibliography}
\end{document}